# Feedback stabilization of multi-qubit Hamiltonian parameters enabled by single-shot measurement-based sequential Monte Carlo

Hyeongyu Jang[1], Jaemin Park[1], Younguk Song[1], Jinwoong Kim[1], Hanseo Sohn[1], Lucas E. A. Stehouwer[2], Davide Degli Esposti[2], Giordano Scappucci[2], and Dohun Kim[1]*

[1]*NextQuantum Center, Department of Physics and Astronomy, and Institute of Applied Physics, Seoul National University, Seoul 08826, Korea*

[2] *QuTech and Kavli Institute of Nanoscience, Delft University of Technology, PO Box 5046, 2600 GA Delft, The Netherlands*

**Corresponding author: dohunkim@snu.ac.kr*

## Abstract

**Fast measurement, signal processing, and accurate estimation of Hamiltonian parameters are essential for feedback control in quantum–classical interface circuitry. However, existing frequentist and Bayesian inference methods typically require a large number of measurements to achieve the accuracy needed to mitigate qubit decoherence. Consequently, feedback control of semiconductor qubits has largely been limited to single-qubit frequency stabilization, whereas two-qubit parameter stabilization remains experimentally unexplored. Here, we demonstrate a real-time feedback framework based on sequential Monte Carlo estimation using one bit of data from a single-shot measurement. Using a four-qubit semiconductor quantum dot device, we rapidly estimate individual qubit frequencies, yielding an approximately twofold increase in coherence time compared with a conventional Bayesian strategy. Moreover, sequential two-qubit parameter estimation using two bits of data enables stabilization of qubit-qubit coupling,**

**allowing both quasi-static frequency drift and exchange-interaction noise to be estimated and suppressed. By shortening the time required for precise parameter estimation, these results demonstrate the importance of the synergistic development of classical and quantum electronics for building robust and scalable quantum technologies in fluctuating environments.**

Real-time feedback control is essential for stabilizing and manipulating complex dynamical systems under noise. In quantum information processing, performing Hamiltonian parameter estimation faster than the characteristic noise timescale is critical for suppressing decoherence and stabilizing universal quantum operations[1–12]. Such estimation-feedback protocols enable rapid interleaved calibration at the hardware layer and prevent qubits operating below the required performance threshold from limiting fault-tolerant operation in large-scale qubit arrays. In these arrays, uniformly stable and well-characterized physical qubits provide the basis for noisy intermediate-scale quantum applications[13] and, ultimately, quantum error correction[14–16].

Unlike classical systems, quantum measurements are intrinsically probabilistic; therefore, repeated single-shot readout is required to obtain reliable expectation values, introducing latency in both data acquisition and processing. As the number of noise sources and the system size increase, extracting noise information for efficient, rapid calibration becomes a scaling bottleneck since simultaneous qubit control[17–19] and readout[20–23] remain practically limited. Prior studies have shown that feedback precision depends on the amount of data used for noise estimation[1,5,24], whereas the associated latency fundamentally limits the achievable feedback rate[8]. In semiconductor spin qubits, feedback-control demonstrations have focused primarily on stabilizing a single-qubit resonance frequency[1,2,4–7], with only a few extensions to two resonance frequencies in two-qubit systems[3,8]. Although Hamiltonian parameter estimation has recently revealed noise features of nearest-neighbor exchange coupling[3,25–27], its feedback performance for active stabilization has not yet been experimentally validated.

The sequential Monte Carlo (SMC) method[28,29], also known as particle filtering, is a state-estimation algorithm that uses observed data and simulated particles in a nonlinear state-space model. SMC-based stochastic filtering has advanced experimental and theoretical research across a range of classical control problems[29–33]. In quantum control, theoretical work has shown that applying SMC to continuously and weakly monitored quantum systems enables the estimation of multiple parameters along the quantum trajectory[34]. By contrast, in the limit of strong projective measurement on a qubit, the outcome is fundamentally discrete and provides only one bit of classical data; therefore, the efficiency of SMC in binary, single-shot measurement regimes remains poorly understood.

Here, we implement a hardware-level Hamiltonian-parameter estimation circuit based on real-time SMC. We demonstrate its control performance in a four-qubit $^{28}$Si/SiGe quantum dot device, in which the qubits are defined by the spin states of confined electrons and coupled through spin-spin exchange interactions for quantum information processing[35]. We demonstrate rapid estimation and feedback control of the resonance frequencies of four qubits and of the exchange coupling between a pair of spins. By leveraging nonlinear stochastic filtering, the implemented recursive estimator requires only one bit of measurement data for single-qubit parameters and two bits for two-qubit parameters to suppress fluctuations under closed-loop operation. Our method improves feedback rates by two orders of magnitude and increases coherence time by approximately two times compared with a conventional Bayesian approach. The proposed control strategy reduces the quantum-measurement overhead required for intermittent calibration and accommodates random noise by incorporating it into the nonlinear state-space model[28,29].

## Device, circuit, and estimation framework

To apply the SMC-based framework to a quantum system, we first describe the device used in the experiments and the circuitry used to implement the real-time estimation protocol. As shown in Fig. 1a, the device consists of a linear four-qubit quantum dot array, Q1–Q4, formed between two charge sensors, SD1 and SD2, in the $^{28}$Si quantum well of a $^{28}$Si/SiGe heterostructure[36] (see Methods). A multilayer gate stack with a 90 nm interdot pitch provides precise tunability of the charge occupation of individual quantum dots and the tunnel couplings between neighboring quantum dots. Charge sensors connected to an LC tank circuit for radiofrequency (RF) reflectometry[37,38] enable detection of the charge occupancy of the quantum dots. By monitoring reflected carrier signals at frequencies of 138 MHz for SD1 and 115 MHz for SD2, we tune the device to the (3, 1, 1, 3) charge configuration, where ($N_1$, $N_2$, $N_3$, $N_4$) denotes the number of electrons in each quantum dot. A detailed description of the experimental setup is provided in the Supplementary Note. 1.

We measure the spin parity of the Q12 and Q34 pairs using Pauli spin blockade[22,39,40] (PSB). For the Q12 pair, ramping the plunger gates from the (3, 1) configuration toward the PSB readout window in the (4, 0) configuration causes the state to collapse to either even ($|11\rangle$, $|00\rangle$) or odd ($|10\rangle$, $|01\rangle$) parity states, (see Extended Data Fig. 1 for the charge-stability diagram and the corresponding data for the Q12 and Q34 pairs). Odd parity yields (4, 0) signals, whereas even parity leads to latching of the (3, 1) signals during a readout time of 4 μs for Q12 (14 μs for Q34). Figure 1b shows that these readout signals can be discriminated with high signal-to-noise ratios of 13.7 and 7.6 for Q12 and Q34, respectively, achieving readout fidelities exceeding 98.8% (see Supplementary Note. 2 for calculations of readout fidelity). For rapid single-qubit operations, we minimize qubit-initialization procedures; the single-shot outcome $m$ is determined adaptively by comparing the spin parity measured before qubit manipulation ($m_i$) with that measured after qubit manipulation ($m_f$), as shown in Fig. 1c. Each single-shot cycle lasts 240 μs, comprising initialization (30 μs), readout (30 μs), compensation pulses to mitigate bias-tee charging effects[41] during initialization and readout (80 μs each), and manipulation (20 μs).

A magnetic field of 0.44 T induces spin-dependent energy splitting, giving a qubit resonance frequency $f_Q$ (Fig. 1d). An integrated micromagnet above the gate stack enables fast electric dipole spin resonance (EDSR) control of individual electron spins[42]. By applying a microwave pulse to the screening gate (Fig. 1a), we acquire EDSR spectra of the four qubits by averaging 200 shots at each microwave frequency, $f_{MW}$ (Fig. 1d). The experimental sequence, including initialization, manipulation through a unitary operation $U$, and measurement of all qubits, is illustrated in Fig. 1e. Although microwave-frequency calibration is conventionally performed using EDSR spectroscopy, which requires more than $10^4$ measurements per qubit, Ramsey experiments can be used for rapid frequency calibration.

For Ramsey interferometry, a unitary operation consisting of two π/2 pulses (X) separated by a free-evolution time $\tau_k$ (Fig. 1f) is applied to the probe circuit shown in Fig. 1e. In the experiment, the number of measurements required to estimate $f_Q$, denoted $N_{shot}$, can be substantially reduced using Bayesian estimation (BE)[24]. In this approach, measurements are restricted to a selected subset of binary outcomes $\{m(\tau_1),\dots m(\tau_k),\dots m(\tau_{100})\}$, acquired at $\tau_k = k \times 80$ ns, as illustrated in Fig. 1f. Nevertheless, interleaved BE routines for the four frequencies $f_{Q1}$, $f_{Q2}$, $f_{Q3}$, and $f_{Q4}$ still require substantial data acquisition time because circuit execution is slow when $N_{shot} = 100$ (Fig. 1e).

In this work, the SMC-based particle filter enables recursive estimation of $f_Q$ from a single-bit Ramsey outcome, $m(\tau_k) = 0$ or 1, thereby reducing $N_{shot}$ from 100 to 1. Below, we describe the four steps of the SMC algorithm, S0–S3, which were experimentally implemented on a field-programmable gate array (FPGA).

**S0.** To estimate an unknown state variable $X_t$ in a quantum system, the particle filter uses sequential importance sampling[28,29,43], in which particles $X_{t-\Delta t}^{(i)}$ ($i = 1 \ldots N_{ptl}$) are drawn from a Gaussian distribution $\mathcal{N}(\mu_{t-\Delta t}, \Sigma_{t-\Delta t})$ with mean $\mu_{t-\Delta t}$ and variance $\Sigma_{t-\Delta t}$ computed from the previous iteration. The initial mean $\mu_0$ is set to the qubit frequencies obtained from the EDSR spectra (Fig. 1d), with $\Sigma_0$ = 200 kHz. Here, $\Delta t$ denotes the estimation cycle of the implemented circuit in Fig. 1e.

**S1.** As illustrated in Fig. 1g, the particles evolve according to Ornstein–Uhlenbeck processes[44] with a transition probability $P(X_t^{(i)} \mid X_{t-\Delta t}^{(i)})$.

**S2.** Once the Ramsey outcome $m(\tau_k)$ is acquired, the $i^{th}$ particle $X_t^{(i)}$ is assigned the weight $w_t^{(i)} = P(m_k \mid X_t^{(i)}) \;/\; \sum_i P(m_k \mid X_t^{(i)})$. The resulting weight distribution approximates the posterior distribution via Monte Carlo integration[28,29,43] $P(X_t \mid m_k) dX_t \sim \sum_i w_t^{(i)} \delta(X_t - X_t^{(i)})$.

**S3.** The distribution of weighted particles yields $f_Q$ by computing the posterior mean $\mu_t$ of the unknown quantum state variable $X_t$ as follows (see Methods for details of the SMC algorithm):

$$\mu_t = \sum_{i=1}^{N_{\text{ptl}}} w_t^{(i)} X_t^{(i)} \quad (1)$$

**Analysis of estimation methods**

To assess the estimation uncertainty of the SMC and BE methods, we first analyzed the closed-loop feedback performance for each qubit and compared the experimental observations with numerical simulations. In closed-loop operation, the control input signal is determined from the frequency error between the estimated $f_Q$ and the target frequency, $f_{\text{target}}$, and is applied to $f_{\text{MW}}$ to suppress fluctuations in $f_Q$ during Ramsey experiments[7].

Figure 2a shows the quasi-static variance, $\sigma^2(N_{\text{shot}}) = 1/(\sqrt{2}\pi T_2^*)^2$, where the coherence time $T_2^*$ is extracted by fitting the Ramsey oscillation to a Gaussian decay. For BE, the initial reduction in $\sigma^2(N_{\text{shot}})$ reflects improved estimation accuracy and approximately follows $\exp(-A \times N_{\text{shot}})$. This exponential behavior is consistent with theoretical work[24], but does not fit well for $N_{\text{shot}} > 100$ because $f_Q$ undergoes diffusive fluctuations over longer estimation times[1,5] at a rate of 280 μs per shot, including a single-shot cycle of 240 μs and a BE calculation time of 40 μs. The simulation incorporating diffusion, as shown by the gray triangles in Fig. 2a, agrees with the observed $\sigma^2(N_{\text{shot}})$, indicating that diffusion limits the estimation accuracy[1,5] (see Supplementary Note. 3). Consequently, the optimum occurs at $N_{\text{shot}} = 80$ for Q1 and $N_{\text{shot}} = 100$ for Q2–Q4.

By contrast, the SMC method using $2^{12}$ particles at $N_{\text{shot}} = 1$, corresponding to an estimation time of 300 μs, exhibits the lowest $\sigma^2(N_{\text{shot}})$ across all datasets and improves the feedback rate by two orders of magnitude compared with BE. To achieve this performance, the SMC protocol requires a sufficiently large number of particles, as shown in Fig. 2b. When only four particles are used, $\sigma^2(N_{\text{ptl}} = 2^2)$ exceeds the $\sigma^2(N_{\text{shot}})$ obtained from BE (Fig. 2a). However, $\sigma^2(N_{ptl})$ decreases rapidly with increasing $N_{ptl}$, and at $N_{ptl} = 2^{12}$, the SMC protocol achieves optimal performance within the available FPGA resources.

Moreover, $\sigma^2(N_{ptl})$ exhibits a characteristic $O(1/N_{ptl})$ scaling, consistent with behavior commonly observed in classical control problems[32,33,45]. To confirm this scaling, we simulate the performance of SMC estimates as in the BE analysis, and the simulation results reproduce the observed $\sigma^2(N_{ptl})$. This agreement suggests that $O(1/N_{ptl})$ is not a system-dependent feature specific to either classical or quantum dynamics, but instead arises from the convergence of approximation schemes toward the true posterior[29,45,46]. Although increasing the number of particles increases the computational cost, this overhead can be mitigated because classical FPGA resources are readily scalable. Notably, the computation time can be substantially reduced because the SMC algorithm is flexible and parallelizable, with independent calculations performed for each particle[29] (see Extended Data Fig. 2). Throughout this work, we use $2^{12}$ particles in the SMC protocol and focus on SMC estimation of single- and two-qubit parameters.

**Single-qubit parameter feedback**

Having characterized the estimation performance for each qubit, we extend the analysis to the simultaneous operation of the four-qubit system. In the probe circuit (Fig. 1e), Ramsey sequences are applied to obtain time traces of the four qubit frequencies. Figure 3a shows simultaneous SMC estimates of $f_Q$ every 960 μs; the corrected frequency traces are acquired using the same probe sequences with adaptive control of $f_{MW}$ (see Methods for details of corrected parameter estimation). When feedback is enabled, all qubit frequencies are simultaneously locked to their respective target frequencies, $f_{target}$, thereby reducing frequency fluctuations and stabilizing the multi-qubit system against noise. The corresponding standard deviations of the time traces $\sigma(\Delta f_Q)$ are reduced by approximately 48.2%, 59.3%, 70.2%, and 58.4% for Q1–Q4, respectively.

Next, we extract the frequency components of the noise data in Fig. 3a using power spectral density (PSD) analysis. In Fig. 3b, all PSDs exhibit overall $1/f$-like behavior, with narrowband peaks at a subharmonic frequency of approximately 5 Hz, possibly due to vibration of the pulse-tube cryocooler[26]. Other harmonic components above 5 Hz, originating from different noise sources, are also visible across a broad frequency bandwidth because of the short estimation cycle of the SMC method. Fitting the feedback-on data to a single Lorentzian function, shown by the black dashed lines in Fig. 3b, yields filter cutoff frequencies $\omega_c$ of 20.8, 17.7, 17.5, and 20.5 rad/s for Q1–Q4, respectively. Quantitatively predicting the effective cutoff frequency remains nontrivial because of the probabilistic and nonlinear structure of the estimation-feedback process. Nevertheless, these $\omega_c$ values exceed the 0.227 rad/s cutoff frequency obtained using BE[7], consistent with the difference in feedback rate. The feedback-off data are not adequately fitted by a single Lorentzian function, indicating that the intrinsic device noise has complex characteristics[26,47]. This result underscores the noise-agnostic nature of the control strategy and its applicability to systems with complex, nontrivial noise profiles.

The noise suppressions shown in Figs. 3a and 3b are also observed in Ramsey traces acquired repeatedly over time. Without feedback control, the Ramsey traces exhibit noisy patterns due to inhomogeneous dephasing (Fig. 3c). By contrast, Fig. 3d shows coherent patterns in the traces, indicating a stabilized oscillation frequency of the Ramsey interference under active frequency feedback. Similar coherent patterns are observed using the BE-based approach (Extended Data Fig. 3); however, compared with SMC, the resulting improvement in coherence time remains modest, as discussed for Figs. 2a and 2b. In Fig. 3e, we summarize the coherence times measured under three conditions: no feedback, BE-based feedback, and SMC-based feedback. Under SMC feedback, the coherence times increase from 6.51, 5.05, 3.78, and 6.54 μs to 9.49, 7.21, 7.13, and 10.05 μs, respectively.

**Two-qubit parameter feedback**

We now consider the SMC feedback control of the two-qubit parameter. In semiconductor spin qubits, the spin-spin exchange interaction is electrically controlled via the barrier gate, which modulates the interdot tunnel coupling (see Supplementary Note. 4). This exchange coupling, $J$, splits the resonance frequency of the target qubit, Q2, depending on the state of the control qubit, Q1, yielding two frequencies, $f_{\mathrm{CROT}} = f_{\mathrm{Q2}} + J/2$ and $f_{\mathrm{zCROT}} = f_{\mathrm{Q2}} - J/2$, corresponding to controlled-rotation (CROT) and zero-CROT (zCROT) operations, respectively[3,25,26]. Therefore, $J$ can be extracted from the frequency difference $f_{\mathrm{CROT}} - f_{\mathrm{zCROT}}$.

Figure 4a shows a $J$-estimation circuit comprising two interleaved Ramsey sequences with different initial qubit states: $|00\rangle$ and $|10\rangle$ to probe $f_{\mathrm{zCROT}}$ and $f_{\mathrm{CROT}}$, respectively. During an $X_{\pi/2}$-wait-$X_{\pi/2}$ Ramsey sequence, $J$ is turned on by a square pulse applied to barrier gate B1 (Fig. 1a). To run this circuit, Q12 is initialized into the deterministic state $|10\rangle$ using the PSB feedback-initialization procedure[22] depicted in Fig. 4b (see Methods for details of qubit initialization). We also attempted deterministic initialization of Q34; however, this was not achieved, possibly due to mixing with other valley states at the anticrossing or device instability. As in single-qubit parameter estimation (Figs. 1–3), each frequency estimation requires only a single measurement, allowing $J$ to be estimated from two bits of data every 1 ms, as shown in Fig. 4c. To suppress noise in the $f_{\mathrm{zCROT}}$ and $f_{\mathrm{CROT}}$ traces, we simultaneously apply active feedback control to the microwave frequency and barrier pulse amplitude using real-time estimates of $f_{\mathrm{zCROT}}$ and $f_{\mathrm{CROT}}$ (see Supplementary Note. 5 for details of the control input).

With feedback control applied to the single- and two-qubit parameters, the standard deviations of the traces decrease from 133.9 to 72.3 kHz for $\sigma(f_{zCROT})$, and from 67.4 to 48.1 kHz for $\sigma(f_{CROT})$. Consequently, $J = f_{CROT} - f_{zCROT}$ is locked to its target coupling of 3.8 MHz, leading to approximately 27% suppression of $\sigma(J)$. The resulting $J$ oscillation (Extended Data Fig. 4) shows an improvement in the oscillation decay time from 2.57 to 3.79 μs. Using the two-qubit parameter traces, we subsequently examine the Pearson correlation coefficient $\rho$ between $f_{zCROT}$ and $f_{CROT}$. The time traces of $f_{zCROT}$ and $f_{CROT}$ obtained without feedback exhibit a positive correlation $\rho = 0.49$, indicating that qubit frequency drift is a significant source of two-qubit gate errors. However, the deviation from perfect correlation, $\rho \neq 1$, suggests the presence of both positively correlated fluctuations and anticorrelated components between the two frequencies. When both the single- and two-qubit parameters are actively stabilized, no appreciable correlation is observed in the corrected traces, with $\rho = -0.03$.

The fluctuations in $f_{zCROT}$ and $f_{CROT}$ observed in the $J$-estimation experiments can lead to phase errors in two-qubit gate operations. Single- and two-qubit parameter corrections are therefore expected to suppress fluctuations in the accumulated conditional phase during the exchange interaction. To verify this, we investigated the controlled-phase (CPhase) oscillation of the blockade probability $P_{blockade}$ as a function of the microwave phase $\phi$, which was used for controlled-NOT (CNOT) gate calibration (Fig. 4d). The calibration procedure involves preparing Q1 in either $|0\rangle$ or $|1\rangle$ and Q2 in a superposition state, as shown in Fig. 4a. After a 100 ns CPhase operation, the phase $\phi$, of the pulse implementing the $Y_\phi$ gate is tuned to realize a CNOT operation. From repeated calibration runs, $n_{run}$, we extract the phase drift $\phi_{drift}$ (Fig. 4e) by fitting each CPhase oscillation to $P_{blockade}(\phi) = a + b\cos(\phi + \phi_{drift})$. The fluctuations in $\phi_{drift}$ from $n_{run} = 1$ to $n_{run} = 25$ are quantified by the standard deviation $\sigma_\phi$. Detailed $\sigma_\phi$ values are provided in the caption of Fig. 4e. To achieve the optimal feedback performance shown in Figs. 4d and 4e, $\sigma_\phi$ is evaluated to optimize the SMC model parameters, namely the diffusion coefficient $D$ and noise correlation time $\tau_c$, as illustrated in Fig. 4f (see Methods for details of the state-space model parameters).

Notably, applying $J$ feedback control to the CPhase gate enables phase-locked operation within a phase uncertainty of 0.07 rad (Fig. 4e), resulting in reliable CNOT operation at $\phi* = 5.46$ rad (Fig. 4d). Accordingly, $\sigma_\phi$ is reduced by approximately 88% and 81% when Q1 is prepared in $|1\rangle$ and $|0\rangle$, respectively. For a Gaussian noise process, the fluctuations measured without feedback control correspond to a phase-noise-limited error of $\epsilon = 0.072$ (see Supplementary Note. 6 for details of the error analysis). With feedback control applied, the error rate is reduced to $\epsilon = 0.002$, corresponding to an upper bound on the fidelity of approximately $1 - \epsilon = 99.8\%$. This error rate includes the phase noise of the arbitrary waveform generator, fitting uncertainty, and estimation errors.

We further evaluated the visibility of the CNOT gate at $\phi* = 5.46$ rad by calculating $(P_{\mathrm{blockade,Q1}|1\rangle} - P_{\mathrm{blockade,Q1}|0\rangle}) \times 100\%$. Figure 4g shows the visibility extracted from each calibration run, yielding a mean visibility of 76.2% $\pm$ 8.9% without feedback, where the error bar denotes $1\sigma$. This result has two important implications. First, the abrupt drop in visibility from 72.4% to 52.3% between the $8^{\mathrm{th}}$ and $9^{\mathrm{th}}$ calibrations highlights that even a calibration performed immediately before the main program does not necessarily ensure reliable CNOT gate performance. Second, as the number of entangled qubits increases, stochastic fluctuations of this magnitude can accumulate and propagate through a sequence of CNOT gates in a large-scale quantum circuit, where ad hoc manual tuning of the microwave phase becomes impractical. These stochastic fluctuations are strongly mitigated under closed-loop conditions. With feedback enabled, the mean visibility increases to 88.7% $\pm$ 0.9%, demonstrating a substantial reduction in stochastic calibration errors, with a nearly one-order-of-magnitude

decrease from 8.9% to 0.9%. Nevertheless, the maximum visibility of 90.4% in the feedback data remains below the phase-limited fidelity of 99.8%, suggesting additional imperfections arising from state-preparation and measurement errors and nonadiabatic errors[48,49], associated with the finite waveform-generator ramp time of approximately 100 ns.

**Conclusion**

Overall, we demonstrated real-time, hardware-level feedback control of semiconductor spin qubits using an estimation framework based on SMC. By enabling rapid parameter tracking with one bit of single-shot data, this approach stabilizes qubit resonance frequencies with estimation times of 300 and 960 μs in the single-qubit and four-qubit experiments, respectively, as well as nearest-neighbor exchange coupling with an estimation time of 1 ms. We confirmed that feedback performance improves with the number of SMC particles, suggesting that further improvement is constrained by available classical hardware resources, unlike conventional Bayesian strategies that rely on repeated quantum measurements with non-negligible overhead. Although we used parallel processing to reduce the computational latency of the SMC algorithm, further reductions should be possible through hardware optimizations, such as lookup tables, pipelining, and reduced algorithmic complexity. Notably, the strong compatibility of semiconductor qubits with complementary metal-oxide-semiconductor fabrication technology, together with a digital very-large-scale integration design ecosystem, offers a tightly integrated platform for this approach, enabling cross-layer design for ultrafast communication and high-bandwidth feedback.

Fast and scalable calibration routines are essential for quantum information processing. Scalable quantum computation motivates applications such as quantum simulation, Shor's factoring algorithm, and variational quantum algorithms; however, realizing these benefits with reduced overhead on practical timescales requires qubit operations with fast intermittent calibration. With a deeper understanding of the underlying noise and control structure, the one-shot SMC protocol will require tailored measurement methods to improve feedback performance beyond the Ramsey-sequence probing strategy used for both BE and SMC in this work. Bridging this gap will require parallel theoretical and experimental efforts to establish a general and useful subroutine for reliable quantum information processing.

# Methods

## Device fabrication

The device was fabricated on an undoped $^{28}$Si/SiGe heterostructure with a $^{29}$Si concentration of 0.08%. A 9 nm tensile-strained quantum well was grown on a strain-relaxed $Si_{0.7}Ge_{0.3}$ substrate approximately 30 nm beneath the surface[36]. Ohmic contacts were formed by phosphorus ion implantation followed by thermal annealing. A 20 nm $Al_2O_3$ gate oxide was deposited on the surface by atomic layer deposition (ALD). Three layers of metallic gates (Ti:Pd = 5:30 nm), separated by 8 nm $Al_2O_3$ oxide layers, were formed subsequently by electron-beam lithography, evaporation, and ALD. A patterned micromagnet (Ti:Co = 15:150 nm) was then deposited by electron-beam evaporation on top of the gate stack, which was terminated by the $Al_2O_3$ oxide layer.

## Qubit initialization

Qubit initialization was performed using two methods tailored to single- and two-qubit gate operations. For single-qubit gate experiments, we used an adaptive initialization scheme to rapidly acquire data for estimating qubit resonance frequencies. With this initialization scheme, we consistently observed high-visibility Rabi oscillations (95.4%–98.2%; Extended Data Fig. 1). PSB feedback initialization was used for two-qubit gate experiments. In this scheme, the spin parity is first measured, and a conditional π rotation, indicated by the double line in Fig. 4b, is applied to Q2 to prepare Q12 in the odd state. This procedure was repeated twice for robust odd-state preparation. In the final PSB readout step, pulsing to the (4, 0) configuration allowed the odd state to relax to the singlet state. Subsequent adiabatic ramping from the (4, 0) to the (3, 1) configuration using the barrier-gate pulse[22,40] initialized Q12 into the $|10\rangle$ state, given the magnetic-field gradient $\Delta B_z$ satisfying $f_{Q1} < f_{Q2}$. Compared with adaptive initialization, PSB feedback initialization doubled the single-shot cycle and reduced the readout fidelity by approximately 2% (see Supplementary Note. 2).

**Nonlinear state-space model**

To implement SMC-based estimation, we formulated the estimation problem in the quantum system as a nonlinear state-space model comprising state and observation models. The state model represents the time-varying dynamics of an unknown state variable, $X_t$, described by a stochastic differential equation,

$$dX_t = \frac{1}{\tau_c}\left(\mu_0 - X_t\right)dt + \sqrt{2D}dW_t \quad (2)$$

where the noise term $W_t$ is a Wiener process, and the diffusion coefficient $D$ characterizes the noise strength. The parameter $\mu_0$ denotes the long-term mean, while the characteristic time $\tau_c$ represents the mean-reversion timescale toward $\mu_0$. $X_t$ follows an Ornstein–Uhlenbeck process, whose transition probability is given by

$$P(X_t \mid X_{t-\Delta t}) = \sqrt{\frac{1}{2\pi D\tau_c(1-e^{-2\Delta t/\tau_c})}}\exp\left[\frac{\left(X_t - X_{t-\Delta t}e^{-\Delta t/\tau_c} - \mu_0(1-e^{-\Delta t/\tau_c})\right)^2}{2D\tau_c(1-e^{-2\Delta t/\tau_c})}\right] \quad (3)$$

Here, $\mu_0$ is set to the qubit frequency obtained from the EDSR spectrum. The parameters $D$ and $\tau_c$ are system-dependent and are determined from experimental data, for example, as shown in Fig. 4f. Supplementary Data Table 2 summarizes the values of $D$ and $\tau_c$ used in the $f_Q$ and $J$ feedback experiments.

The observation model relates the state variable to noisy observations. In Ramsey experiments, the single-shot measurement process used to obtain a binary outcome, $m(\tau_k)$ = {0, 1}, is modeled by a conditional probability distribution for the weight update

$$P(m(\tau_k)\,|\,X_t) = \frac{1}{2}\left(1 + (-1)^{m(\tau_k)}\{\alpha + \beta\cos(2\pi(f_{\mathrm{MW}} - X_t)\tau_k + \theta)\}\right) \quad (4)$$

where $\alpha$ ($\beta$) is a parameter determined by the rotation axis (visibility) on the Bloch sphere, and $\theta$ denotes the initial phase of the Ramsey oscillation. Because the unknown state variable $X_t$ corresponds to the qubit resonance frequency, the microwave frequency $f_{\mathrm{MW}}$ is determined from the oscillation frequency of the Ramsey fringe. For all estimations of $f_Q$, $f_{\mathrm{zCROT}}$, and $f_{\mathrm{CROT}}$, measurement outcomes were obtained from 0.5 MHz Ramsey oscillations.

This estimation procedure remains valid when feedback control is applied. After each probe step, $f_{\mathrm{MW}}$ is adaptively updated based on the probe data within a latency of 1 μs. As a result, subsequent estimation of the unknown state variable $X_t$ is performed under closed-loop conditions in which frequency fluctuations, $f_{\mathrm{MW}} - X_t$, are effectively suppressed, as shown in Fig. 3a. In Fig. 4c, this procedure is applied, together with additional control of the pulse amplitude (see Supplementary Note. 5 for details of the control input).

**SMC algorithm**

Based on the nonlinear state-space model described above, the algorithm used in the main text is as follows.

1. Draw particles $X_{t-\Delta t}^{(i)}$ from $\mathcal{N}(\mu_{t-\Delta t}, \Sigma_{t-\Delta t})$

2. For $X_{t-\Delta t}^{(i)}$ ($i = 1, \ldots, N_{\text{ptl}}$), obtain $X_t^{(i)}$ from the transition probability

$$P(X_t \mid X_{t-\Delta t}^{(i)}) = \sqrt{\frac{1}{2\pi D\tau_{\text{c}}(1-e^{-2\Delta t/\tau_{\text{c}}})}} \exp\left[\frac{\left(X_t - X_{t-\Delta t}^{(i)} e^{-\Delta t/\tau_{\text{c}}} - \mu_0(1-e^{-\Delta t/\tau_{\text{c}}})\right)^2}{2D\tau_{\text{c}}(1-e^{-2\Delta t/\tau_{\text{c}}})}\right]$$

3. Calculate the weight of $X_t^{(i)}$ by

$$w_t^{(i)} = \frac{1}{2}\left(1+(-1)^{m(\tau_k)}\{\alpha+\beta\cos(2\pi(f_{\text{MW}} - X_t^{(i)})\tau_k + \theta)\}\right)$$

4. Normalize the weights as $w_t^{(i)} / \sum_{i=1}^{N_{\text{ptl}}} w_t^{(i)}$

5. Estimate the mean and variance by

$$\mu_t = \sum_{i=1}^{N_{\text{ptl}}} w_t^{(i)} X_t^{(i)}$$
$$\Sigma_t = \sum_{i=1}^{N_{\text{ptl}}} w_t^{(i)} (X_t^{(i)} - \mu_t)^2$$

6. Return to step 1 for the next estimation with $\mu_t$ and $\Sigma_t$

## Data availability

The data that support the findings of this study are available from the corresponding author upon request.

## Acknowledgments

This work was supported by a National Research Foundation of Korea (NRF) grant funded by the Korean Government (Ministry of Science and ICT (MSIT)) (RS-2023-00283291), RS-

2024-00413957, SRC Center for Quantum Coherence in Condensed Matter RS-2023-00207732, RS-2023-NR077112 and Quantum Technology R&D Leading Program (Quantum Computing) RS-2024-00442994) and a core center program grant funded by the Ministry of Education (No. 2021R1A6C101B418). Lucas E. A. Stehouwer and Davide Degli Esposti developed and characterized the heterostructure under the supervision of Giordano Scappucci. Correspondence and requests for materials should be addressed to DK (dohunkim@snu.ac.kr).

## Author contributions

J.P. fabricated the device. H.J. performed the measurements. J.P., Y.S., and H.S. built the experimental setup and configured the measurement software. J.K. contributed to the development of the estimation algorithm. J.P., Y.S., J.K., and H.S. contributed to data analysis. L.E.A.S., D.D.E., and G.S. synthesized and provided the $^{28}$Si/SiGe heterostructure. H.J. and D.K. conceived the project, discussed the results, and wrote the manuscript with the contributions of all authors.

## Competing interests

The authors declare no competing interests.

## Figures

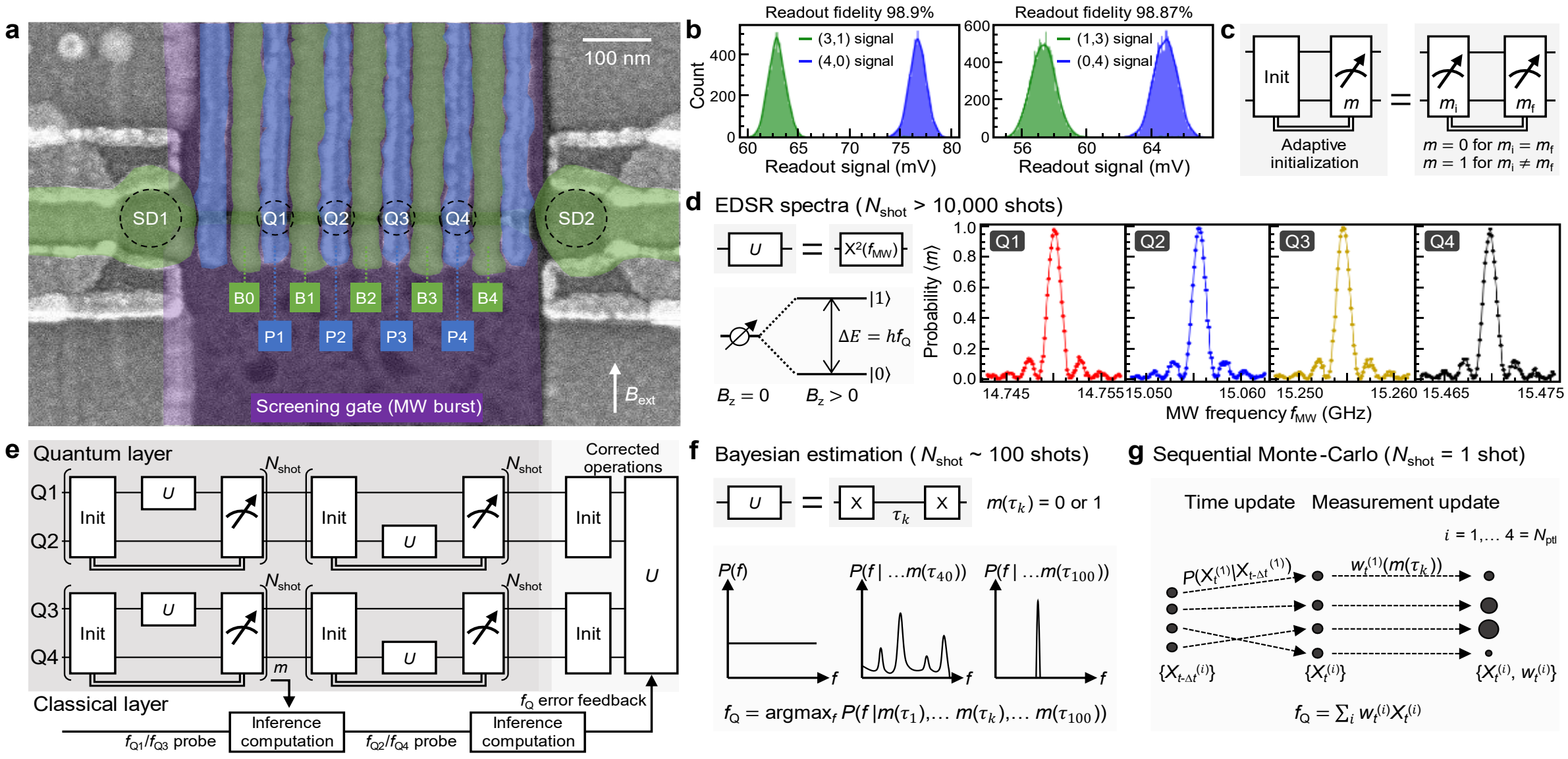

**Figure 1. Device, circuit, and estimation framework. a.** False-colored scanning electron microscopy image of the $^{28}$Si/SiGe device. Purple, green, and blue denote the first, second, and third metallic gate layers, respectively. Quantum dots located between sensor dots SD1 and SD2 are defined by screening gates (purple), plunger gates (blue, P1–P4), and barrier gates (green, B0–B4). **b.** Histogram of radiofrequency-demodulated readout signals for Q12 (left panel) and Q34 (right panel). **c.** Schematic blocks of the adaptive initialization protocol. **d.** Top left: electric dipole spin resonance (EDSR) pulse indicating an $X_\pi$ rotation ($X^2$). Bottom left: Zeeman splitting induced by the magnetic field $B_z$. Right: EDSR spectra obtained using adaptive initialization. **e.** Experimental sequence comprising initialization (Init), manipulation ($U$), and readout, repeated $N_{\mathrm{shot}}$ times to estimate the qubit frequencies $f_Q$, which are subsequently used to correct single-qubit operations. **f.** Top: Ramsey pulse sequence used for Bayesian estimation (BE). Bottom: schematic showing the BE update from the uniform distribution $P(f)$ to the posterior distribution $P(f|\ldots m(\tau_{100}))$. Measurement outcomes $m(\tau_k)$ are used to update the distribution according to Bayes' theorem. $f_Q$ is determined by the value $f^*$ that maximizes $P(f|\ldots m(\tau_{100}))$. **g.** Schematic of the sequential Monte Carlo (SMC) algorithm showing time and measurement updates. $N_{\mathrm{ptl}}$ denotes the total number of particles.

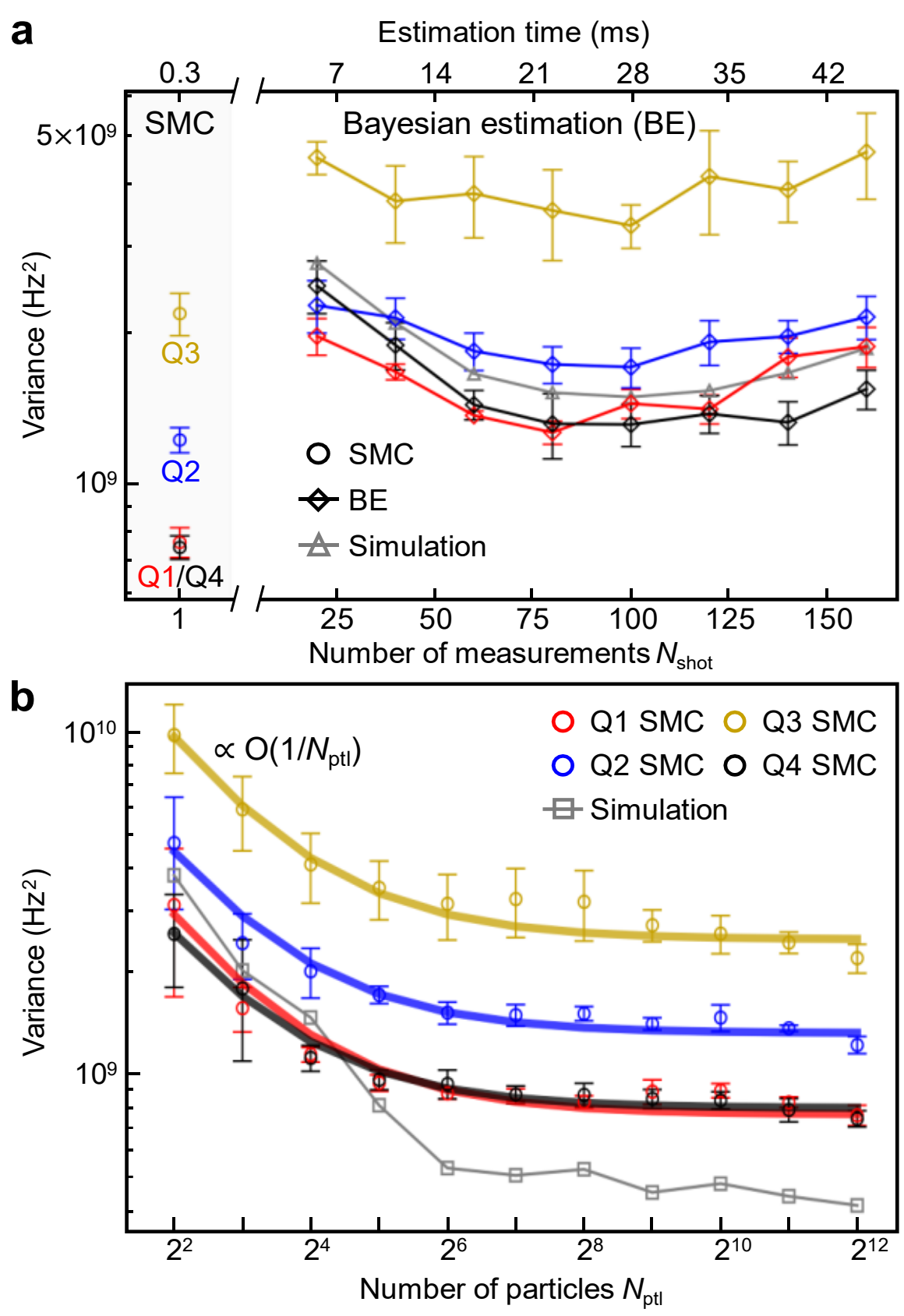

**Figure 2. Analysis of estimation methods. a.** Quasi-static variances for each of the four qubits as a function of the number of measurements and the estimation time. The data are fitted to $\exp(-A \times N_{shot}) \times B$, yielding $A$ = 0.008, 0.005, 0.004, and 0.012 and $B$ = 2,302, 2,539, 4,669, and 3,102 kHz$^2$ for Q1–Q4, respectively. The fitted curves are not shown for clarity. Gray triangles indicate the mean squared error (MSE) obtained from the simulation using BE. **b.** Quasi-static variances as a function of $N_{ptl}$ in the SMC particle filter. Solid curves show fits to $C_0 / N_{ptl} + C_1$. The fitted values are $C_0$ = 8,715, 12,700, 29,080, and 7,186 kHz$^2$, $C_1$ = 767, 1,326, 2,491, and 801 kHz$^2$ for Q1–Q4, respectively. Gray squares indicate the MSE obtained from the simulation using SMC. Error bars denote 1σ.

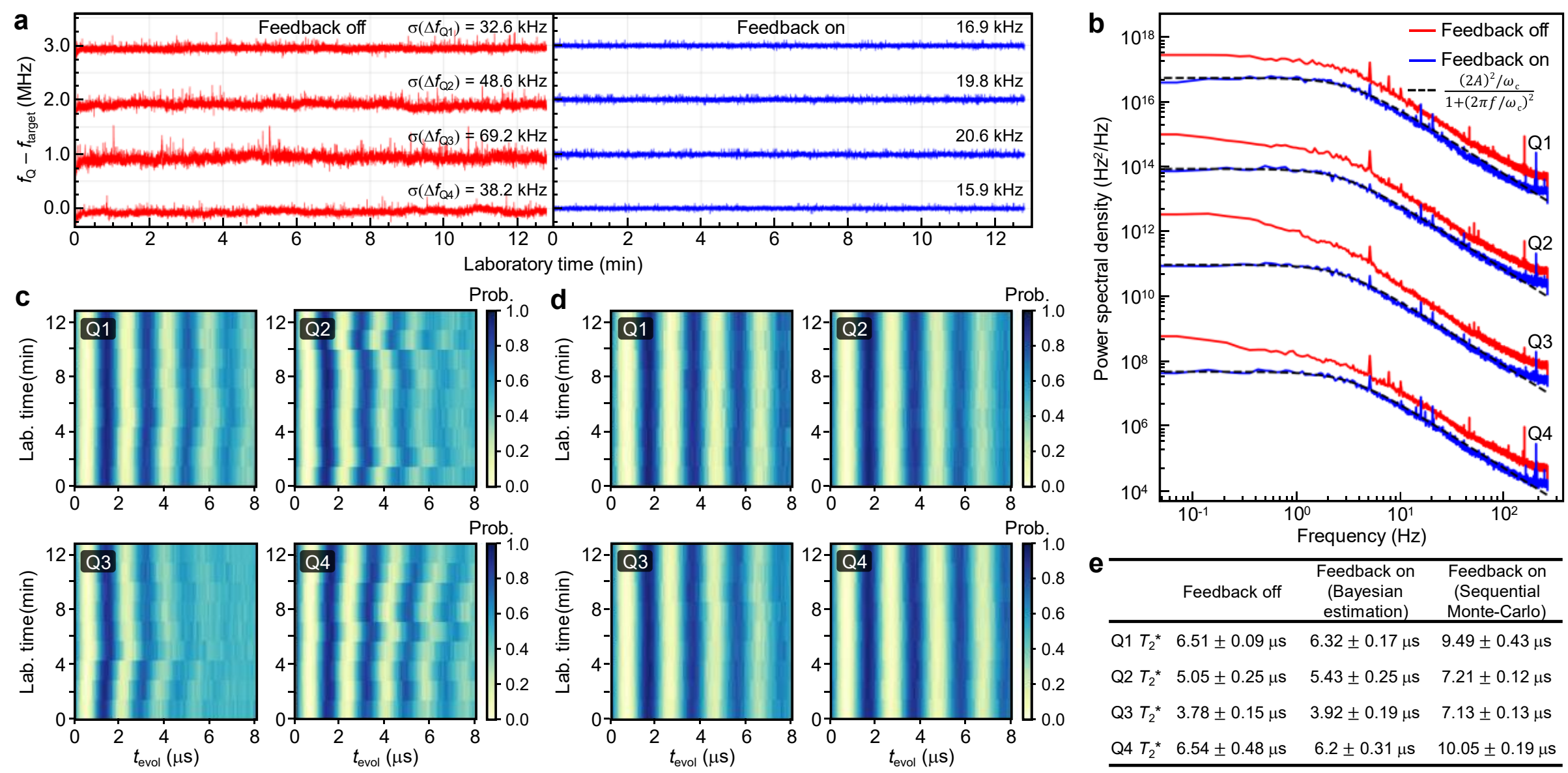

**Figure 3. Qubit-frequency feedback. a.** Example time traces of the estimated qubit frequencies without (left) and with (right) adaptive control of the microwave frequency. For clarity, the traces are offset from their respective target frequencies, $f_{\text{target}}$, by 0 MHz (Q4), 1 MHz (Q3), 2 MHz (Q2), and 3 MHz (Q1). The standard deviation of each trace is indicated above the corresponding trace. Here, $\Delta f_Q = f_Q - f_{\text{target}}$. **b.** Power spectral density (PSD) of the traces shown in **a**. Black dashed lines represent fits to a Lorentzian function $S(f) = ((2A)^2 / \omega_c) / (1 + (2\pi f / \omega_c))$. The extracted amplitudes are $A$ = 17.2, 19.8, 20.7, and 15.9 kHz for Q1–Q4, respectively. The cutoff frequency $\omega_c$ is provided in the main text. For clarity, the PSDs are vertically offset by multiplicative factors of 1 (Q4), $10^3$ (Q3), $10^6$ (Q2), and $10^9$ (Q1). **c, d.** Repeated Ramsey interference as a function of the free-evolution time $t_{\text{evol}}$ measured without (**c**) and with (**d**) adaptive control of the microwave frequency. **e.** Coherence times, $T_2$*, measured with no feedback, BE-based feedback, and SMC-based feedback. Error bars denote 1σ. All data in panels **a**–**e** were obtained during simultaneous four-qubit operations.

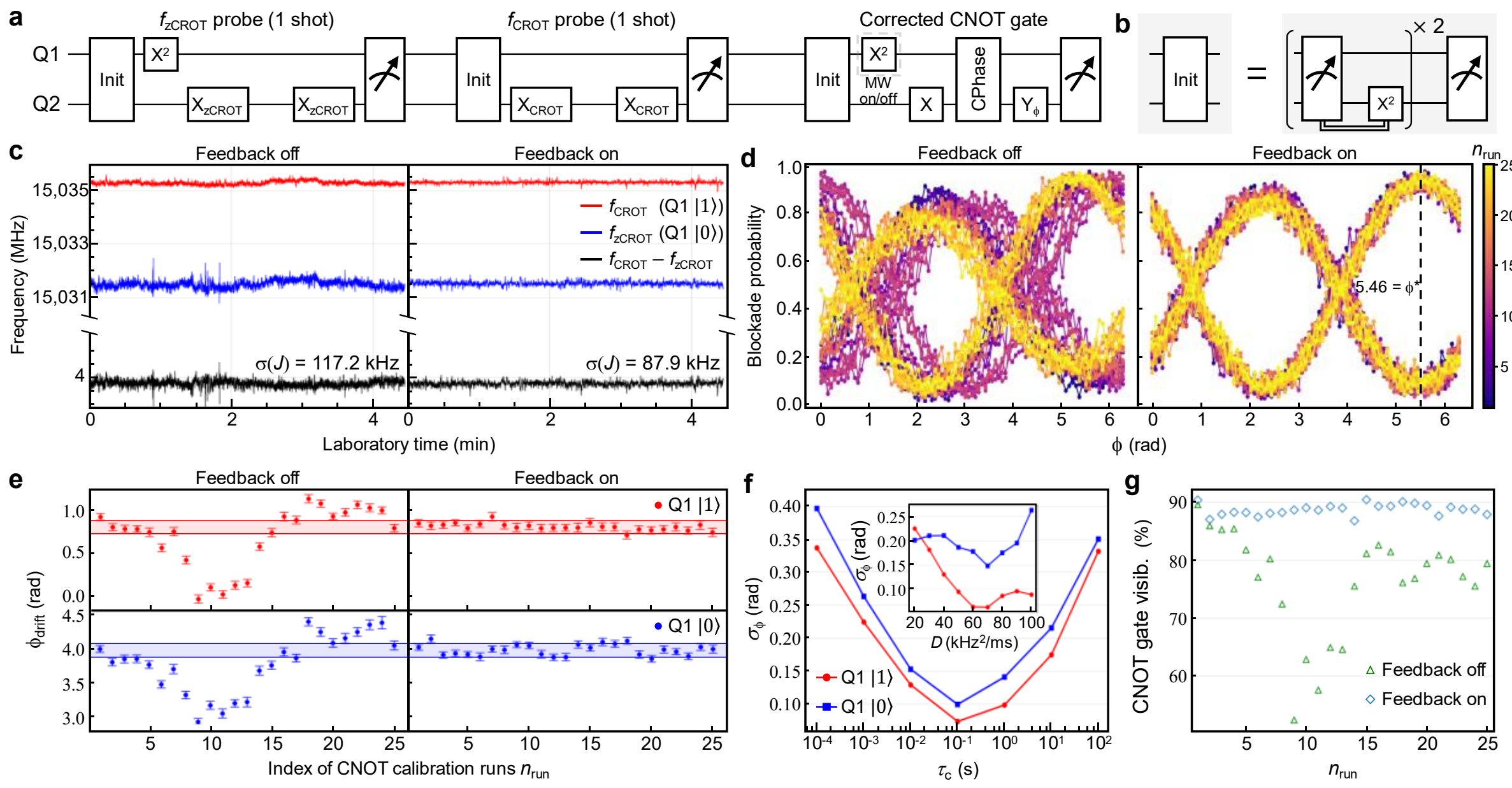

**Figure 4. Exchange-coupling feedback. a.** Experimental sequence comprising the zero-controlled-rotation frequency ($f_{\mathrm{zCROT}}$) probe, controlled-rotation frequency ($f_{\mathrm{CROT}}$) probe, and corrected controlled-NOT (CNOT) operation used for the experiments shown in panels **d**–**g**. The gray dashed box indicates an optional $\pi$ pulse. **b.** Schematic blocks representing the feedback-initialization protocol. **c.** Example time traces of the estimated $f_{\mathrm{zCROT}}$ and $f_{\mathrm{CROT}}$, with the exchange coupling defined as $J = f_{\mathrm{CROT}} - f_{\mathrm{zCROT}}$. The interleaved sequence for $f_{\mathrm{zCROT}}$ and $f_{\mathrm{CROT}}$ is performed without (left) and with (right) simultaneous adaptive control of the microwave frequency and barrier-pulse amplitude. **d.** Consecutively measured controlled-phase oscillations as a function of the microwave phase $\phi$. Each color represents an individual calibration run, $n_{\mathrm{run}}$. **e.** Phase drift $\phi_{\mathrm{drift}}$ extracted from panel **d**, with error bars representing $3\sigma$ from the fit. The shaded region indicates a reference phase uncertainty of 0.07 rad. The fluctuations represented by the standard deviation are $\sigma_\phi = 0.35$ rad (top left), 0.42 rad (bottom left), 0.04 rad (top right), and 0.08 rad (bottom right). **f.** Dependence of $\sigma_\phi$ on the noise correlation time, $\tau_c$, and diffusion coefficient, $D$ (inset). **g.** Visibility extracted from each calibration run at $\phi^* = 5.46$ rad in **d**.

## Extended data figures

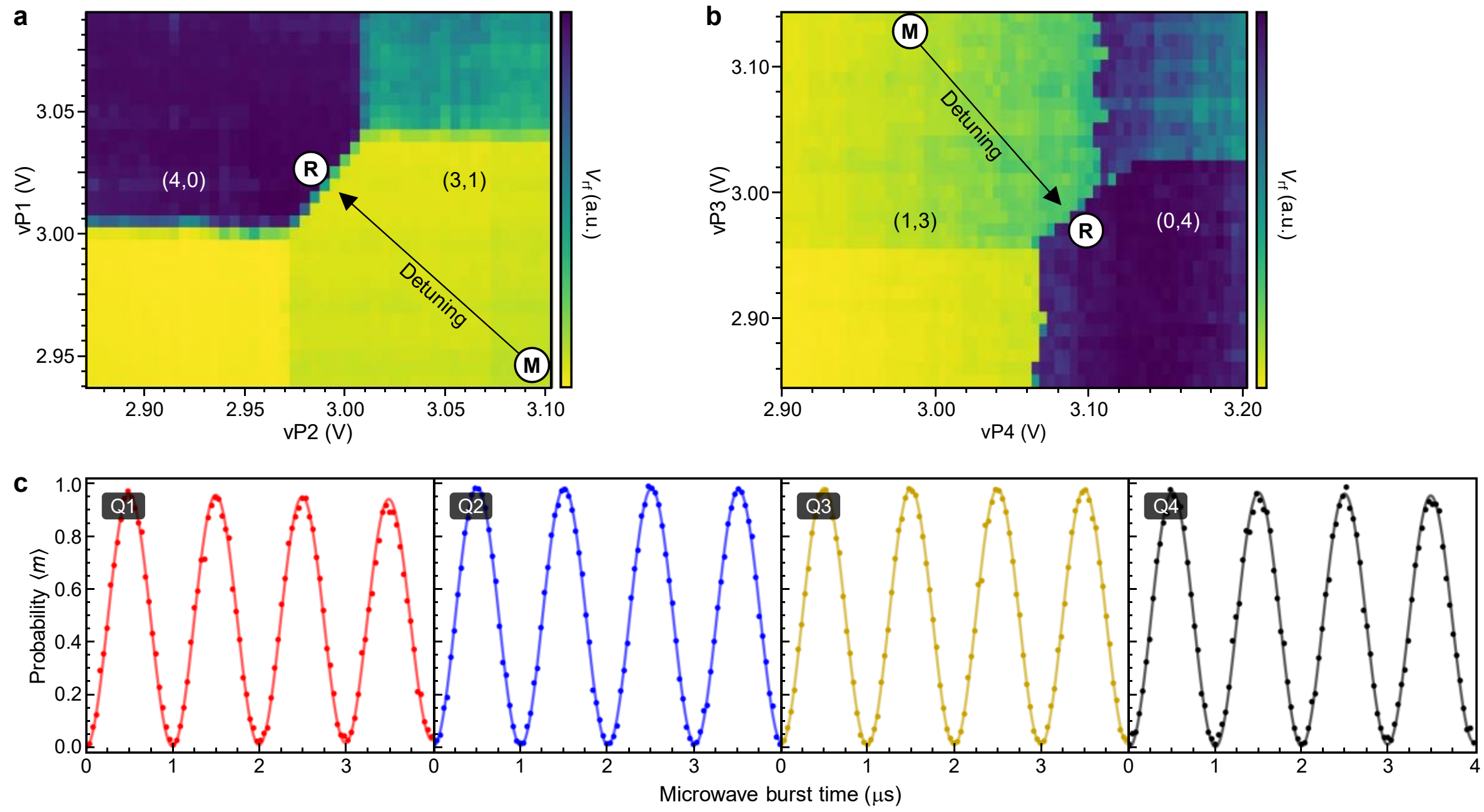

**Extended Data Figure 1. Charge stability diagram and Rabi oscillations. a, b.** Charge stability diagram taken at the (4,0)/(3,1) anticrossing for Q12 (**a**) and the (1,3)/(0,4) anticrossing for Q34 (**b**). The point indicated with 'M' indicates the qubit manipulation point, and the point indicated as 'R' indicates the readout point. For the Q34 pair, ramping the plunger gates vP3 and vP4 from the (1, 3) configuration toward the PSB readout window in the (0, 4) configuration causes the state to collapse to either even parity or odd parity states. Odd parity yields (0, 4) signals, whereas even parity leads to latching of the (1, 3) signals during a readout time of 14 μs. **c.** Rabi oscillations as a function of the microwave burst time for every qubit. The data are acquired using adaptive initialization. Visibilities are 95.4, 98.2, 97.6, and 96.4% for Q1-Q4, respectively.

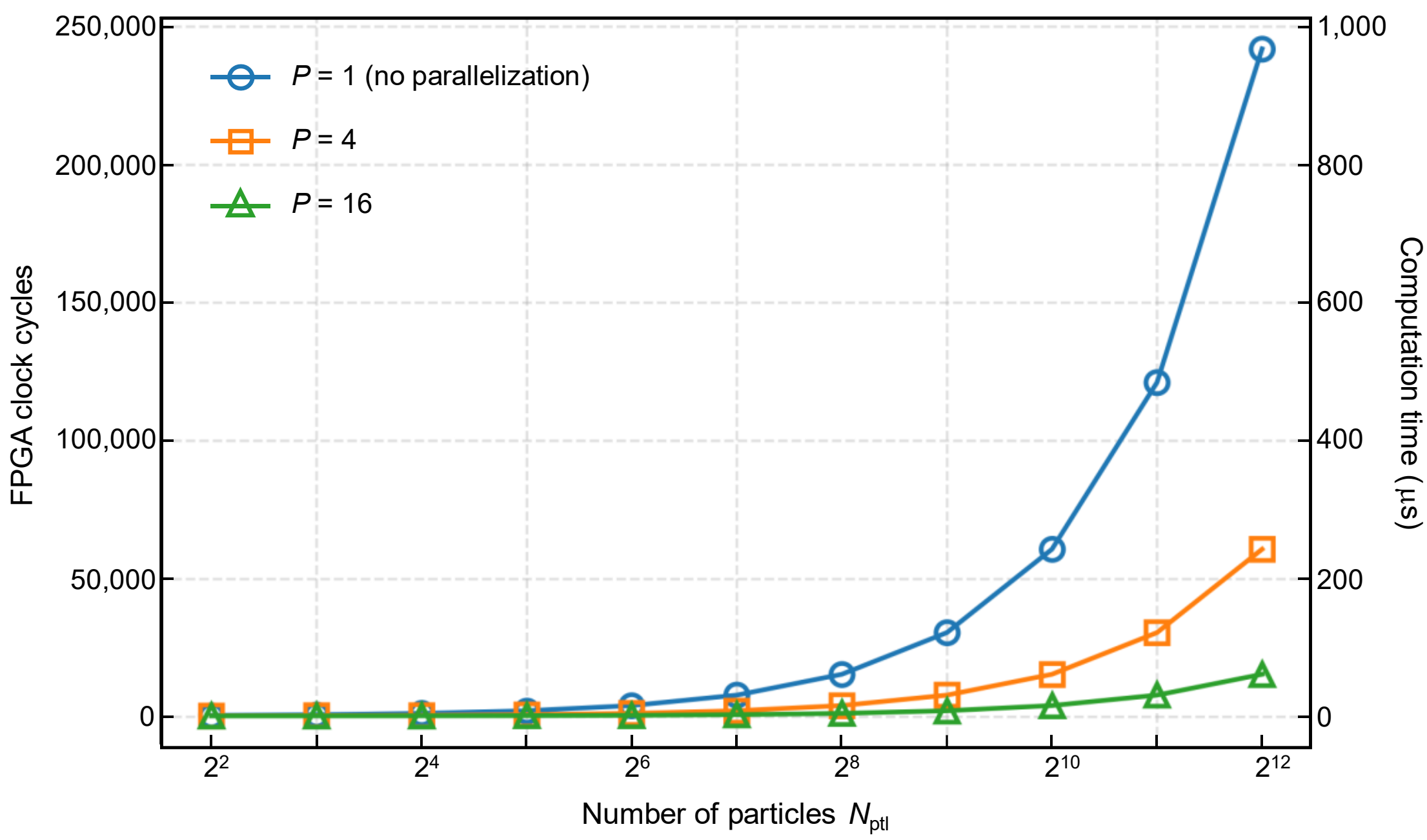

**Extended Data Figure 2. Scaling of FPGA clock cycles with particle number and parallelization.** Clock cycles increase with the number of particles $N_{\text{ptl}}$, while increasing the degree of parallelization, $P$, reduces the computation time of the SMC algorithm, up to the limit set by available resources. Solid lines show clock cycles = 209 + 236 × $N_{\text{ptl}}$ / $P$ in the implemented hardware (Quantum Machines OPX+).

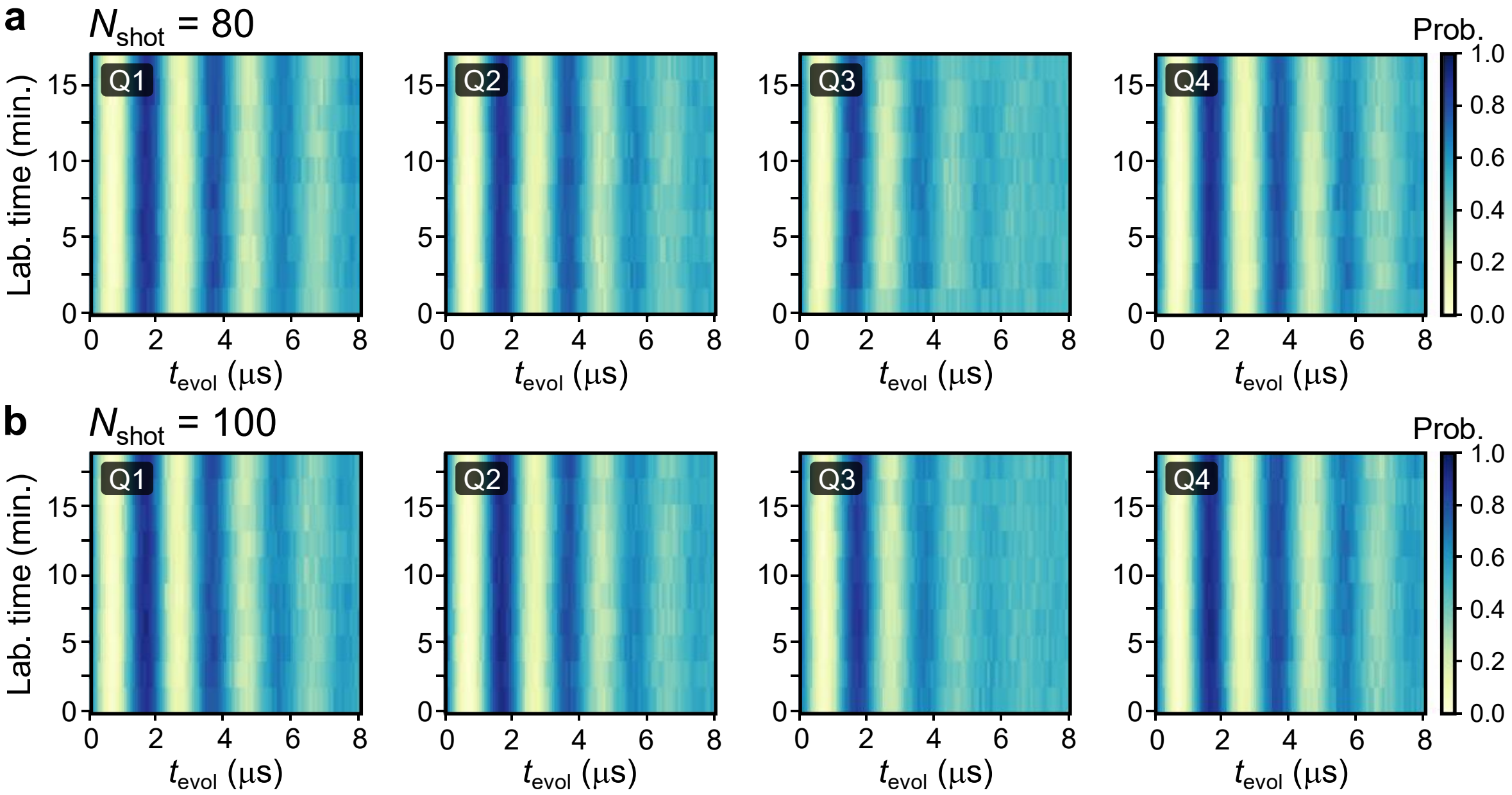

**Extended Data Figure 3. Ramsey traces using Bayesian estimation. a, b.** Repeated Ramsey interference as a function of the free-evolution time $t_{evol}$. Bayesian estimation is performed at $N_{shot}$ = 80 (**a**) and $N_{shot}$ = 100 (**b**) with adaptive control of the microwave frequency.

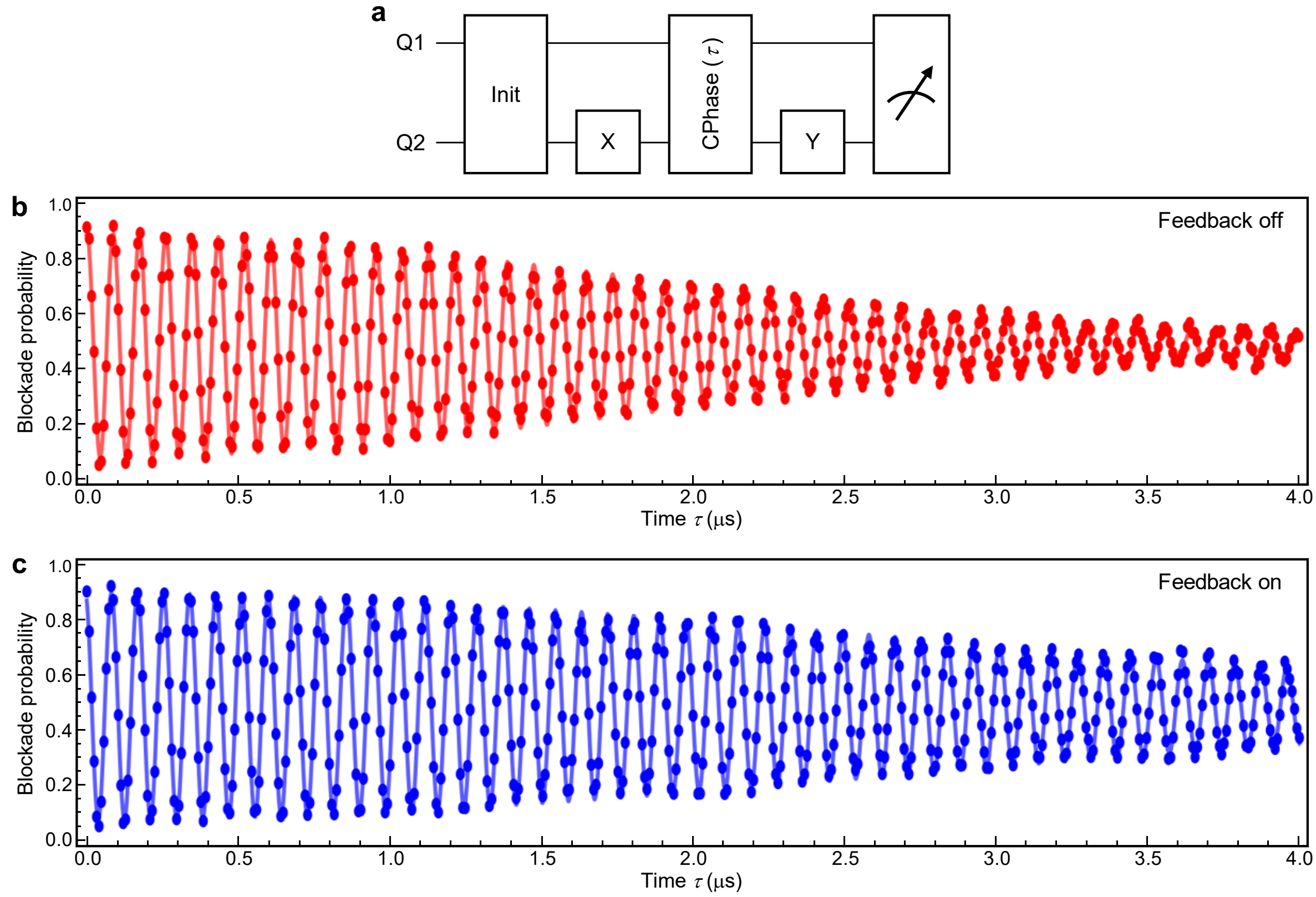

**Extended Data Figure 4. *J*-oscillation experiment. a.** Experimental sequence used to measure $J$ oscillations. **b, c.** $J$ oscillation as a function of the evolution time $\tau$ without (**b**) and with (**c**) feedback control. The measured quality factors, defined as $Q = f \times T$, are 29.6 (**b**) and 44 (**c**). These values are obtained from the decay time of the evolution $T = 2.57$ μs and the oscillation frequency $f = 11.53$ MHz for panel **b**, and $T = 3.79$ μs and $f = 11.6$ MHz for panel **c**.

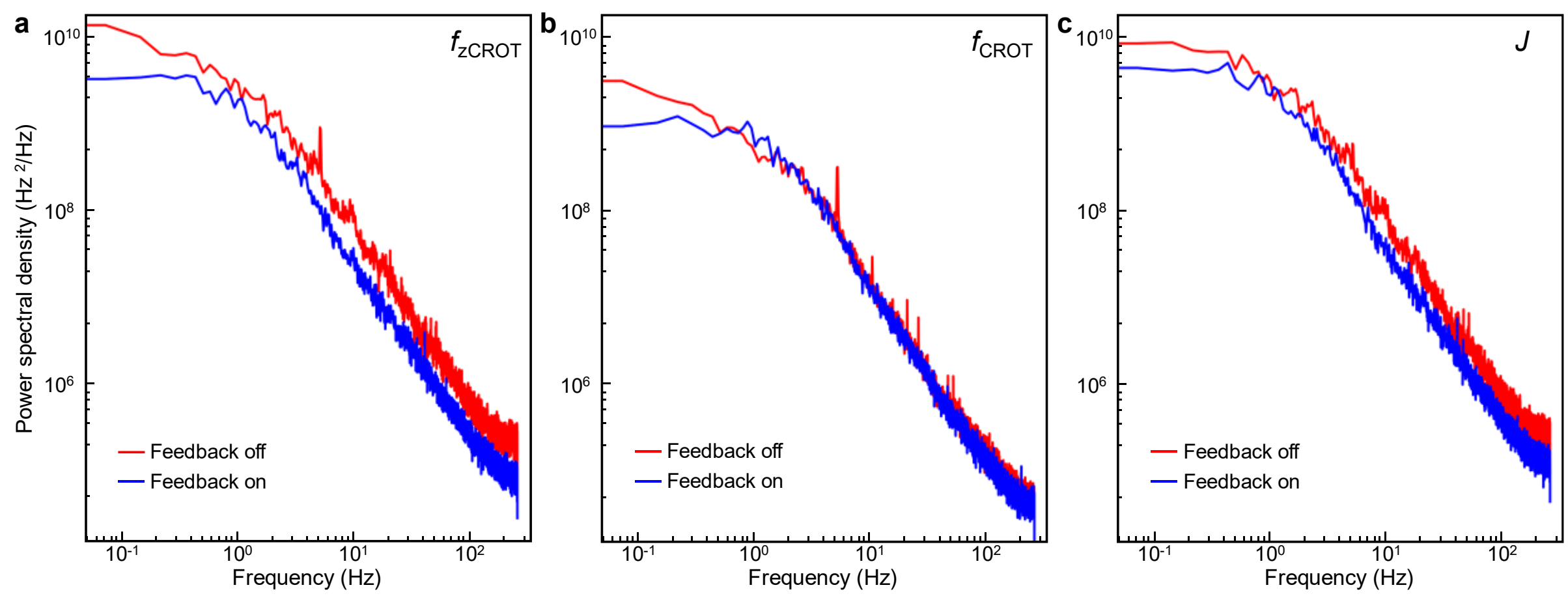

**Extended Data Figure 5. Two-qubit parameters in the frequency domain. a-c.** PSDs of the traces for $f_{\mathrm{zCROT}}$ (**a**), $f_{\mathrm{CROT}}$ (**b**), and $J$ (**c**), shown in Fig. 4c. Fitting the feedback-on data to a single Lorentzian function yields noise amplitudes $A$ = 72.2, 47.3, and 86.5 kHz, and filter cutoff frequencies $\omega_{\mathrm{c}}$ = 6.0, 7.6, and 6.1 rad/s for $f_{\mathrm{zCROT}}$, $f_{\mathrm{CROT}}$, and $J$, respectively.

# Feedback stabilization of multi-qubit Hamiltonian parameters enabled by single-shot measurement-based sequential Monte Carlo

Hyeongyu Jang[1], Jaemin Park[1], Younguk Song[1], Jinwoong Kim[1], Hanseo Sohn[1], Lucas E. A. Stehouwer[2], Davide Degli Esposti[2], Giordano Scappucci[2], and Dohun Kim[1]*

[1] *NextQuantum Center, Department of Physics and Astronomy, and Institute of Applied Physics, Seoul National University, Seoul 08826, Korea*

[2] *QuTech and Kavli Institute of Nanoscience, Delft University of Technology, PO Box 5046, 2600 GA Delft, The Netherlands*

*Corresponding author: dohunkim@snu.ac.kr*

**Supplementary Information**

**Supplementary Note 1. Experimental setup**

The device was operated in a cryogen-free dilution refrigerator (Oxford Instruments Triton-500) with a base temperature of approximately 8 mK. The metal gates were connected to low-noise dc voltage sources (Stanford Research Systems, SIM928), with selected gates additionally receiving baseband and microwave signals through on-board bias tees. The baseband control and I/Q-modulated microwave pulses were generated by an arbitrary waveform generator (Quantum Machines OPX+ and SDT QCU) and by a signal generator equipped with I/Q mixers and local oscillators (Quantum Machines Octave). The RF carrier signals at − 46 dBm were generated using a lock-in amplifier (Zurich Instruments, UHFLI) and attenuated through the cryogenic attenuators and the directional coupler by − 60 dB. The reflected carrier signals, modulated by the conductance of the charge-sensor dot, were amplified by 45 dB with cryogenic low-noise amplifiers at the 4 K stage (Cosmic Microwave

Technology CITLF2, two stages in series) and further amplified by 25 dB at room temperature (Lotus Communication Systems LNA100M2P0G). The amplified signals were demodulated through UHFLI and then recorded by the quantum controller (OPX+). Overall pulse sequencing and real-time data processing were coordinated using the Quantum Universal Assembly (QUA) framework.

**Supplementary Note 2. PSB parity readout error**

We evaluate the fidelity of the PSB parity readout[1–4] using the following matrix,

$$\begin{pmatrix} F_{00} & 1-F_{11}(F_{01}) \\ 1-F_{00}(F_{10}) & F_{11} \end{pmatrix} \qquad (1)$$

$F_{00}$ ($F_{11}$) denotes the probability of measuring odd (even) parity when the qubits are prepared in the odd (even) state. The off-diagonal elements represent errors of the state preparation and measurement process. From this matrix, we obtain the average readout fidelity $F$ = $(F_{00}+F_{11})/2$. Throughout this paper, we present the raw measurement data and do not remove the readout error. In the table below, we summarize the elements of each matrix.

| | Q12 | Q34 | Q12 |
|---|---|---|---|
| | Initialized into the odd state | | Initialized into $\lvert 10\rangle$ |
| $F_{00}$ | 0.9978 | 0.997 | 0.9894 |
| $F_{10}$ | 0.0022 | 0.003 | 0.0106 |
| $F_{01}$ | 0.0198 | 0.0196 | 0.0448 |
| $F_{11}$ | 0.9802 | 0.9804 | 0.9552 |
| $F$ | 0.989 | 0.9887 | 0.9723 |

**Supplementary Data Table 1. Readout matrix elements.** The vB1 barrier pulse with an amplitude $A_{B1}$ = 120 mV, together with vP1 and vP2 detuning pulses (a 200 ns ramp time commonly), is used to prepare Q12 into $\lvert 10\rangle$ ($A_{B1}$ = 0 for odd-state preparation) when ramping back to the operating point in the (3,1) charge configuration (see Methods and Ref.[3] for details of pulse sequences in the feedback-initialization scheme). We speculate that the vB1 pulse used for adiabatic $\mathrm{S}-|10\rangle$ transition also causes adiabatic $\mathrm{S}-\mathrm{T}_{-}$ transition[4], increasing the initialization errors in $F_{10}$ and $F_{01}$.

**Supplementary Note 3. Simulation of estimation protocols for numerical convergence analysis**

We simulate the performance of Bayesian estimation and sequential Monte Carlo. We assume that the simulated noise is quasi-static in the single measurement of 280 μs, but varies between measurements according to the diffusion rate[5–7]. For Bayesian estimation, we simulate Ramsey outcomes for $N_{\mathrm{shot}}$ different measurements using the random variable $X \in [0, 1]$, following the Binomial distribution $P(X=k)=\binom{1}{k}p^k(1-p)^{1-k}$ with probability of the Ramsey oscillation $p$. Then, we use the simulated outcomes to compute the estimate of $f_{\mathrm{Q}}$ in the single-qubit Hamiltonian $H=\frac{hf_{\mathrm{Q}}}{2}\sigma_z$, where $h$ is Planck's constant, and $\sigma_z$ is the Pauli-Z matrix. In sequential Monte Carlo, the estimate is computed from a single simulated outcome ($N_{\mathrm{shot}}$ = 1). By repeating this procedure 50,000 times for both estimation protocols, we then extract the mean squared error, MSE = $\mathbb{E}[(f_{\mathrm{Q}} - f_{\mathrm{Q}}^{\mathrm{estimated}})^2]$, as a function of $N_{\mathrm{shot}}$ or $N_{\mathrm{ptl}}$.

One plausible explanation for the $1/N_{\mathrm{ptl}}$ convergence of sequential Monte Carlo shown in Fig. 2b is the strong law of large numbers. If we simulate $N_{\mathrm{ptl}}$ independent and identically distributed random samples $X_t^{(i)}$ according to an unknown distribution $P(X_t)$, an empirical estimate of this distribution is given by a normalized histogram $P(X_t) \approx \frac{1}{N_{\mathrm{ptl}}}\sum_{i=1}^{N_{\mathrm{ptl}}}\delta_{X_t^{(i)}}(X_t - X_t^{(i)})$, where $\delta_{X_t^{(i)}}$ denotes the Dirac-delta mass. We further obtain the empirical estimate of the following function $I_N(f_t) = \frac{1}{N_{\text{ptl}}}\sum_{i=1}^{N_{\text{ptl}}} f(X_t^{(i)})$, corresponding to $I(f_t) = \int f(X_t)P(X_t)dX_t$. The variance of $I_N(f_t)$ is described by[8]

$$Var\left[\frac{1}{N_{\text{ptl}}}\sum_{i=1}^{N_{\text{ptl}}} f(X_t^{(i)})\right]=\left(\frac{1}{N_{\text{ptl}}}\right)^2\sum_{i=1}^{N_{\text{ptl}}} Var\left[f(X_t^{(i)})\right]=\left(\frac{1}{N_{\text{ptl}}}\right)^2\sum_{i=1}^{N_{\text{ptl}}}\sigma_f^2 \qquad (2)$$

$\sigma_f^2$ is equal to the variance of $I(f_t)$ from the strong law of large numbers, yielding $Var[I_N] = \sigma_f^2 / N_{\text{ptl}}$ and in a similar way $\mathbb{E}[I_N] = I$. Consequently, the rate of convergence is $1 / N_{\text{ptl}}$, as $N_{\text{ptl}} \rightarrow \infty$, for the estimation error, MSE = $\mathbb{E}[(I - I_N)^2] = \mathbb{E}[(I_N - \mathbb{E}[I_N])^2] = Var[I_N] = \sigma_f^2 / N_{\text{ptl}}$.

Here, the unknown distribution $P(X_t)$ represents the true posterior, with the posterior variance $I(f_t) = \int (X_t - \langle X_t \rangle)^2 P(X_t)dX_t$ or the posterior mean $I(f_t) = \int X_t P(X_t)dX_t$. Unlike this example, however, drawing random samples directly from this unknown distribution is impossible in practice; particle filtering approximates the posterior distribution by a large set of Dirac-delta masses (particles), which evolve over time according to the system dynamics and incoming observation. A more detailed proof for the convergence of particle filtering is available in the probability theory literature[9,10].

**Supplementary Note 4. Two-qubit Hamiltonian parameters under exchange interaction**

The Hamiltonian of Q1-Q2 qubits can be described by[11]

$$H = \frac{hf_{Q1}}{2}\sigma_z^{(1)} \otimes I^{(2)} + \frac{hf_{Q2}}{2} I^{(1)} \otimes \sigma_z^{(2)} + \frac{hJ}{4}\left(\sigma^{(1)} \bullet \sigma^{(2)} - I^{(1)} \otimes I^{(2)}\right) \qquad (3)$$

where $f_{Q1}$ and $f_{Q2}$ denote the spin resonance frequencies of qubits 1 and 2, $h$ Planck's constant, $I^{(k)}$ the identity matrix, and $\sigma^{(k)}$ is the Pauli matrix acting on qubit $k$. $J$ represents the exchange strength described by the Heisenberg model[12]. In the presence of the magnetic field $B_z$, we define the single-qubit resonance frequencies by $hf_{Q1} = g\mu_B B_{z,1}$ and $hf_{Q2} = g\mu_B B_{z,2}$ where $\mu_B$ is the Bohr's magneton and $g \approx 2$ is the g-factor in silicon. The Hamiltonian in the basis of $|11\rangle$, $|\widetilde{10}\rangle$, $|\widetilde{01}\rangle$, and $|00\rangle$ is given by[6,13]

$$H = \frac{h}{2}\begin{pmatrix} \Sigma & 0 & 0 & 0 \\ 0 & -\sqrt{\Delta^2 + J^2} - J & 0 & 0 \\ 0 & 0 & \sqrt{\Delta^2 + J^2} - J & 0 \\ 0 & 0 & 0 & -\Sigma \end{pmatrix} \qquad (4)$$

Here, $\Sigma = f_{Q2} + f_{Q1}$, $\Delta = f_{Q2} - f_{Q1}$, and the tilde indicates the hybridization of the spin eigenstates $|10\rangle$ and $|01\rangle$ due to the exchange coupling. The exchange coupling $J$ splits the resonance frequency of one qubit, enabling conditional rotation of one qubit dependent on the state of the other qubit. The corresponding transition frequency of Q2 can be expressed as

$$\begin{aligned} f_{\mathrm{CROT}} &= \frac{E_{11} - E_{10}}{h} \approx f_{Q2} + \frac{J}{2} + \frac{J^2}{4|\Delta|} \\ f_{\mathrm{zCROT}} &= \frac{E_{01} - E_{00}}{h} \approx f_{Q2} - \frac{J}{2} + \frac{J^2}{4|\Delta|} \end{aligned} \qquad (5)$$

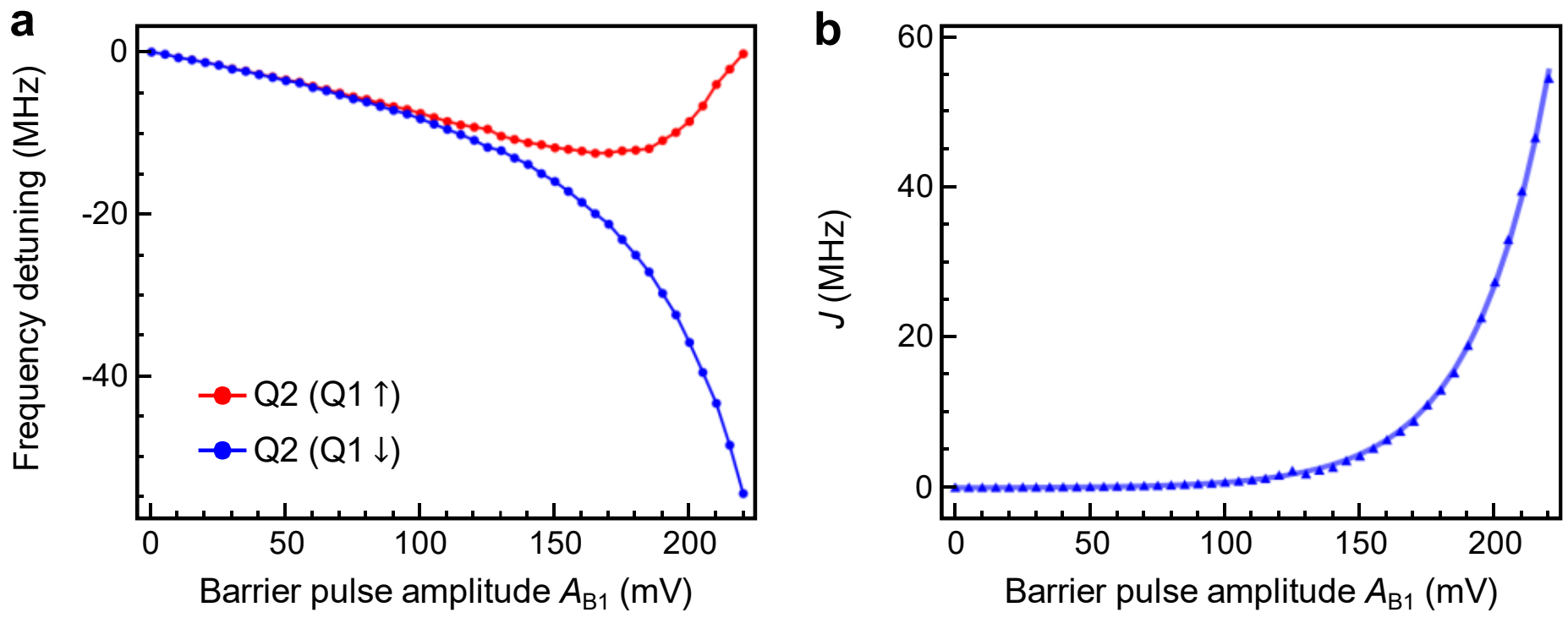

**Supplementary Figure 1. Engineering of exchange interaction. a.** Frequency detuning of the target qubit Q2 conditional on the state of the control qubit Q1 as a function of barrier pulse amplitude $A_{B1}$. The horizontal axis represents the voltage applied to the B1 gate. **b.** Exchange strength $J$ as a function of barrier pulse amplitude. The data are extracted from **a**.

## Supplementary Note 5. Converting qubit parameter estimates into feedback control inputs

For closed-loop operation of single- and two-qubit parameters, the microwave frequency $f_{\mathrm{MW}}(t)$ and the B1 pulse amplitude $A_{\mathrm{B1}}(t)$ are updated using control input signals $u(t)$,

$$\begin{aligned} f_{\mathrm{MW}}(t) &= f_{\mathrm{target}} - u_f(t) \\ A_{\mathrm{B1}}(t) &= A_{\mathrm{target}} - u_A(t) \end{aligned} \qquad (6)$$

where $f_{\mathrm{target}}$ and $A_{\mathrm{target}}$ are constant, and $u(t) \neq 0$ ($u(t) = 0$) when feedback is on (off). To determine input signals, we consider a noise model of $f_{\mathrm{CROT}}$ and $f_{\mathrm{zCROT}}$ described by

$$\begin{aligned} f_{\mathrm{CROT}} &= f_{\mathrm{target}} + \frac{J_{\mathrm{target}}}{2} + \delta f + \frac{df_{\mathrm{CROT}}}{dA_{\mathrm{B1}}} \times \delta V \\ f_{\mathrm{zCROT}} &= f_{\mathrm{target}} - \frac{J_{\mathrm{target}}}{2} + \delta f + \frac{df_{\mathrm{zCROT}}}{dA_{\mathrm{B1}}} \times \delta V \end{aligned} \qquad (7)$$

Here, $\delta f$ represents noise-induced deviations from the target resonance frequency $f_{\mathrm{target}}$, $\delta V$ denotes fluctuations in the barrier gate, and the gradients $\frac{df_{\mathrm{CROT}}}{dA_{\mathrm{B1}}}$ and $\frac{df_{\mathrm{zCROT}}}{dA_{\mathrm{B1}}}$ are extracted from Supplementary Fig. 1a. By solving for $\delta f$ and $\delta V$ using the real-time estimates of $f_{\mathrm{CROT}}$ and $f_{\mathrm{zCROT}}$ in Eqn. (7), we determine $u_f(t) = -\delta f$ and $u_A(t) = \delta V$,

$$u_A = (f_{\mathrm{CROT}} - f_{\mathrm{zCROT}} - J_{\mathrm{target}}) \Big/ \left( \frac{df_{\mathrm{CROT}}}{dA_{\mathrm{B1}}} - \frac{df_{\mathrm{zCROT}}}{dA_{\mathrm{B1}}} \right)$$
$$u_f = -\left( \frac{f_{\mathrm{CROT}} + f_{\mathrm{zCROT}} - 2 f_{\mathrm{target}}}{2} - \frac{u_A(t)}{2} \left( \frac{df_{\mathrm{CROT}}}{dA_{\mathrm{B1}}} + \frac{df_{\mathrm{zCROT}}}{dA_{\mathrm{B1}}} \right) \right) \tag{8}$$

When exchange coupling is turned off ($f_{\mathrm{Q}} \approx f_{\mathrm{CROT}} \approx f_{\mathrm{zCROT}}$, $J_{\mathrm{target}} = 0$), the control law in Eqn. (8) remains valid for single-qubit frequency feedback, with $u_A(t) = 0$ and $u_f(t) = -\left( f_{\mathrm{Q}} - f_{\mathrm{target}} \right)$.

**Supplementary Note 6. Phase fluctuation-dominated two-qubit error analysis**

In this section, we describe the error caused by fluctuations in the controlled-phase (CPhase) oscillation. In the presence of noise in $f_{\mathrm{CROT}}$ and $f_{\mathrm{zCROT}}$, the target qubit acquires a random phase offset $\phi$ during a 100 ns CPhase operation, which reduces the probability contrast when averaged over these fluctuations. The resulting contrast is therefore given by the ensemble average $1 - \epsilon_\alpha = \mathbb{E}[\cos \phi]$, where $\phi$ is a Gaussian random variable with the standard deviation $\sigma_\phi$, $\epsilon_\alpha$ denotes the error rate, and $\alpha$ represents the state of the control qubit. Under this assumption, the expectation value can be evaluated analytically, yielding $\epsilon_\alpha = 1 - \exp(-\sigma_\phi^2 / 2)$. Using the measured $\sigma_\phi$ (Fig. 4e), we obtain the average error rate $\epsilon = (\epsilon_1 + \epsilon_0)/2$.

## Supplementary Note 7. State-space model parameter

| Single- and two-qubit parameters | Model parameters | |
|---|---|---|
| | $D$ (kHz$^2$/ms) | $\tau_c$ (ms) |
| $f_{Q1}$ | 60 | 150 |
| $f_{Q2}$ | 60 | 150 |
| $f_{Q3}$ | 60 | 150 |
| $f_{Q4}$ | 60 | 150 |
| $f_{CROT}$ | 70 | 100 |
| $f_{zCROT}$ | 75 | 100 |

**Supplementary Data Table 2. Model parameters used to estimate each qubit frequency.**

## Supplementary References